# Resistivity and Thermoelectric power of $Na_xCoO_2$ (x =1.0, 0.7 and 0.6) system

H. Kishan, V.P.S. Awana[$,*], M.A. Ansari, Anurag Gupta and R.B. Saxena

National Physical Laboratory K.S. Krishnan Marg, New Delhi 110012, India

V. Ganesan and A.V. Narlikar

Inter-University Consortium for DAE Facilities, University Campus, Khandwa Road, Indore-452017, MP, India

C. A. Cardoso[*]

Center for Superconducity Research, University of Maryland, College Park, MD 20742-4111, USA

R. Nirmala, Devendra Buddhikot and S.K. Malik

Tata Institute of Fundamental Research Homi Bhabha Road, Mumbai 400005, India

Results of thermo-electric power (S) and electrical resistivity ($\rho$) measurements are reported on $Na_xCoO_2$ compounds with x = 1.0, 0.7 and 0.6. These are single-phase compounds crystallizing in the hexagonal structure (space group $P6_3$/mmc) at room temperature. Thermo-electric power values at 300K ($S_{300K}$) are, $\cong 80\mu V/K$, $39\mu V/K$ and $37\mu V/K$ for x = 1.0, 0.7 and 0.6 samples, respectively. The samples with x=0.7 and 1.0 are metallic down to 5 K, while the x = 0.6 sample is semiconducting. The value of $\rho_{300K}$ for x = 1.0 sample is ~0.895 m$\Omega$-cm and the power factor ($S^2/\rho$) is ~ 7.04 x $10^{-3}$ W/mK$^2$ which qualifies it as a good thermo-electric material. In x =1.0 sample, S(T) is positive throughout 300–5K temperature range and decreases monotonically to zero as temperature $T \rightarrow 0$. In contrast, S(T) of x = 0.7 and 0.6 samples changes sign and shows negative values between 90 K and 16 K before approaching zero as $T \rightarrow 0$. Anomalous S(T) behavior of x = 0.6 and 0.7 samples, which are coincidentally the precursor materials to the reported superconductivity in this class of materials, indicates a dramatic change in the electronic structure of these compounds on lowering the Na content.

**INTRODUCTION**

In recent years, the compound $NaCoO_2$ has been studied extensively due to its good thermoelectric response (see, for instance, refs. 1 and 2). For a promising thermoelectric material, the essential requirements are high thermoelectric power (S) and low resistivity ($\rho$) and low thermal conductivity ($\kappa$) [3]. In fact, materials with figure of merit $Z = S^2/\rho\kappa$, reaching a value close to unity are considered good thermoelectric materials. Though oxides are thought to be very poor thermoelectric materials, primarily due to their low electrical conductivity, the compound $NaCoO_2$ proved it otherwise [1,2]. It is believed that, in this compound, good electrical conduction is ensured by $CoO_2$ block and high thermoelectric power is contributed from the Na-O phonon-drag layer [1,2]. Keeping in view that the thermal conductivity ($\kappa$) for $NaCoO_2$ remains in the range of 10-15 mW/cmK over temperature range of 5 to 400 K [1,2], one can focus more on the S and $\rho$. In this situation the shortened name for figure of merit $Z = S^2/\rho\kappa$ is termed as power factor ($S^2/\rho$). In the present article we will be discussing mainly the power factor ($S^2/\rho$) value for the studied samples.

The magnetic structure of the compound is reported to be complex and the exact arrangement of the Co spins is yet unclear. In particular, the nature of magnetic transition below 20 K is yet unknown [4,5]. Physical properties of $NaCoO_2$ change dramatically with Na content in the final product [1-6]. Very recently, $Na_xCoO_2$ became the focus of interest due to the appearance of superconductivity with de-intercalation of Na and inclusion of water in the parent structure, resulting in $Na_{0.35}CoO_2:1.3H_2O$ with a superconducting transition temperature of around 4.5 K [7]. It seems that the Na content is important in deciding not only the thermoelectric properties but also the appearance of superconductivity.

We report here transport properties of $Na_xCoO_2$ samples with x = 1.00, 0.70 and 0.60. Worth mentioning is the fact that the heat treatment schedule followed by us [see below] gives excellent agreement between nominal starting and end values of Na content. The thermoelectric properties of x = 1.00 sample are comparable to those reported in the literature [1,4]. Further, we find that for x values of 0.70 and 0.60, an anomalous behavior of S as a function of temperature is seen in terms of a crossover of its sign at around 90 K and before going to the absolute zero value at 5 K. This behavior suggests drastic changes in electronic structure of the compound with x before appearance of reported superconductivity in further lowering Na content and $H_2O$ interclated compound $Na_{0.35}CoO_2:1.3H_2O$.

**EXPERIMENTAL DETAILS**

The samples of the $Na_xCoO_2$ series with x=1.0, 0.7 and 0.6 were synthesized by a solid-state reaction route using $NaCO_3$ and $Co_2O_3$ in stoichiometric proportions. The heating procedure was precisely the same as used by Motohashi *et al* [6] i.e. "rapid heat-up" technique in which the mixed powders were put directly in a pre-heated furnace at $750^0C$ and fired for 18 hours. The fired powder was thoroughly ground and pressed into a circular pellet. The pellet was finally fired at $850^0C$ for 24 hours, before cooling it slowly to room temperature over a span of 6 hours. X-ray diffraction (XRD) patterns were obtained at room temperature (MAC Science: MXP18VAHF; Cu$K_\alpha$ radiation). Resistivity measurements were carried out by the conventional four-probe method (PPMS, Quantum Design). Thermoelectric power (TEP) measurements were carried out by dc differential technique over a temperature range of 5 – 300 K, using a home made set up. Temperature gradient of ~1 K was maintained throughout the TEP measurements.

**RESULTS AND DISCUSSION**

The room temperature X-ray diffraction (XRD) patterns of $Na_xCoO_2$ samples with x = 1.00, 0.70 and 0.6 are shown in Fig. 1. All the samples are reasonably single phase crystallizing in the hexagonal λ-$Na_xCoO_2$ structure [4-6] [space group P$6_3$/mmc], though few $Co_3O_4$ impurity lines are seen close to the background level. The lattice parameters are a = 2.823 (3) Å and c = 10.943 (5) Å for x = 1.00, a = 2.821 (3) Å and c = 10.948 (7) Å for x =0.70 and a = 2.820 (2) Å and c = 10.949 (8) Å for x =0.60. These lattice parameters are in agreement with those previously reported [4-6]. The resistivity (ρ) versus temperature (T) plots for x = 1.00 and 0.70 samples in zero magnetic field and in 6T applied field are shown in Fig. 2. Both samples exhibit metallic behavior down to 2 K with room temperature resistivity ($\rho_{300K}$) of nearly 0.89 mΩ-cm and 6.50 mΩ-cm for x = 1.00 and x = 0.70 samples, respectively. Though both x = 1.00 and 0.70 samples exhibit qualitatively the metallic type of conduction, quantitatively there are couple of finer interesting differences. For example, the x = 1.00 sample exhibits constant dρ/dT of around 2.74μV/K in the temperature range of 50 – 3000 K. Below 50 K the dρ/dT value is decreased, demonstrating changed conduction process below this temperature. For x = 0.70 sample the dρ/dT is comparatively less but positive above 270 K to that below this temperature until down to 2 K, where it is is nearly 2.7μV/K. For x = 0.60 sample, the $\rho_{RT}$ is found to be high (82 mΩ-cm) and it shows semiconducting nature of electrical conduction down to 2 K with ρ (5 K) of 645 mΩ-cm (plot not shown). Recently the phase diagram of $Na_xCoO_2$ was explored, and it was shown that charge ordering of $Co^{2+/3+}$ occurs at x = 0.50, giving rise to nearly an insulating electrical conduction in the system [8]. Our x = 0.60 sample is close to the $Co^{2+/3+}$ charge ordered composition of $Na_xCoO_2$ and hence that could be responsible for highly semiconducting behavior. Another interesting fact is that applied magnetic field up to 6 Tesla has very little effect on the general resistivity behaviour of all the samples. This is in general agreement with earlier reports [2,4-6], and is reminiscent of the fact that Co spins do not order magnetically with long range in studied temperature range. As far as x = 0.60 sample is concerned its $Co^{2+/3+}$ ordering does not melt in applied field of 6 Tesla and hence no change in conduction process under applied magnetic field.

The main result of the present letter is depicted in Fig. 3, i.e. the thermoelectric power (S) versus temperature (T) plots for $Na_xCoO_2$ compound with x = 1.00, 0.70 and 0.60. Room temperature thermoelectric power ($S_{300K}$) decreases with increase in x. For x = 1.00 sample the $S_{300K}$ value is around ≅ 80 µV/K, and decreases nearly to half of this value for x = 0.70 sample. For x = 0.60 sample, the $S_{300K}$ value is further decreased to ≅ 35 µV/K. This is consistent with earlier reports that $S_{300K}$ value decreases with decreasing x below 1.00. As far as x = 1.00 sample is concerned, its $S_{300K}$ (≅ 80 µV/K) and $ρ_{300K}$ (0.89 mΩ-cm) values are good enough for the material to be considered as a thermoelectric material, with further scope of optimization [1-6]. S decreases slowly, having small +ve slope with T for x = 1.00 sample down to say 100 K, and later decreases with relatively higher slope until around 16 K, below which S decrement is again small positive with linear decrement. Finally at 5 K, the S reaches a value close to zero. This trend of S vs. T plot for x = 1.00 sample is in general agreement with earlier reports. Further seen is a kink (marked by $T^*$) in S-T plot at around 240 K. This kink in S-T or ρ-T plots is seen generally in $Na_xCoO_2$ compounds and is attached to second order structural/phase transition [1,6].

For x = 0.70 and 0.60 samples S decreases with T with a constant slope and passes from +ve S to −ve S at around 80 K. Further below 70 K the value S increase from −ve to absolute zero with decrease in T. The x = 0.70 sample finally reaches zero S at 5 K and joins the S-T plot for x = 1.00 sample. The S-T plot for x = 0.60 sample is similar to that as for x = 0.70. A kink is further seen in S-T plot for this sample at same temperature as for x = 1.00 sample, which is supposedly having the same origin as given for x = 1.00 sample.

Here the point of discussion is that though several reports [1-6] exist on S-T measurements for $Na_xCoO_2$, the kind of plot given in current article for x = 0.70 and 0.60 is not seen earlier. Interestingly most of the earlier work either pertained to the optimization of thermoelectric properties of the compound with higher x values (x > 0.70) [1-6] or the realization of superconductivity (x < 0.35) [7]. The kind of S-T plot seen in our present article indicates towards complex electronic changes in this material before the superconductivity sets in at further lower x values. The situation is comparable to high temperature superconducting cuprates, where a purely magnetic compound turns to a spin glass system before the superconductivity sets in. We do not have as such any concrete explanation for the S-T plots seen for x = 0.70 and 0.60 samples and this needs further more work.

In conclusion, for $Na_xCoO_2$ compounds the S(T) and ρ(T) results are obtained with a good figure of merit ($S^2/ρ$) for x = 1.00 sample. Further, an anomalous S(T) is seen for x = 0.70 and 0.60 samples in comparison to x = 1.00.

**FIGURE CAPTIONS**

Figure 1. X-ray diffraction patterns of $Na_xCoO_2$ (x = 1.00 and 0.70) samples.

Figure 2. Resistivity ($\rho$) vs. temperature (T) for $NaCoO_2$ and $Na_{0.7}CoO_2$ in 0 and 6 Tesla applied fields.

Figure 3. Thermoelectric power (S) vs. temperature (T) for $Na_xCoO_2$ with x = 1.00, 0.70 and 0.60. $T^*$ is the S(T) kink temperature seen for x = 1.00 and 0.6 samples. Solid line shows the linear S(T) below 16 K for x = 1.00 sample.

Fig.1 H. Kishan et al.

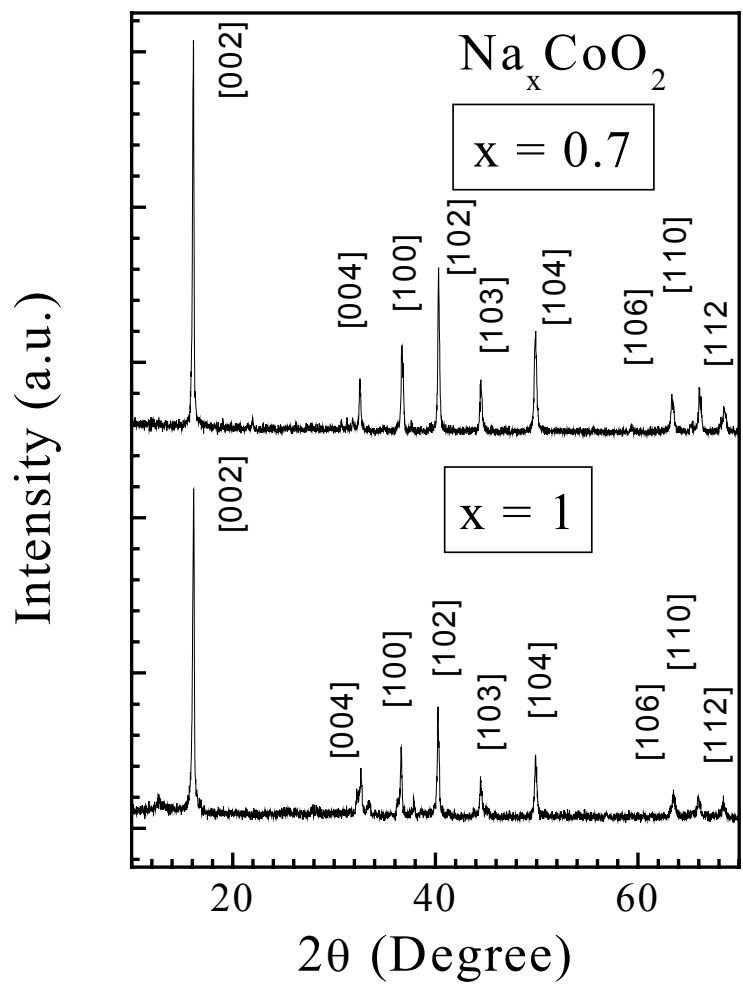

Fig.2 H. Kishan et al.

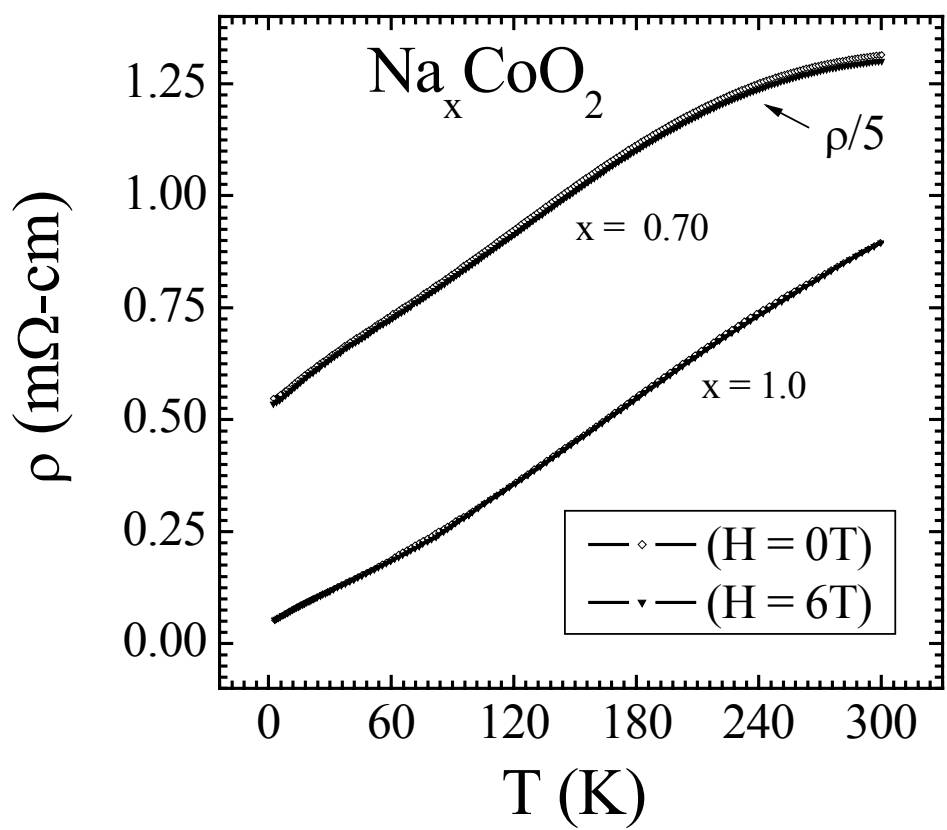

Fig. 3 H. Kishan et al.

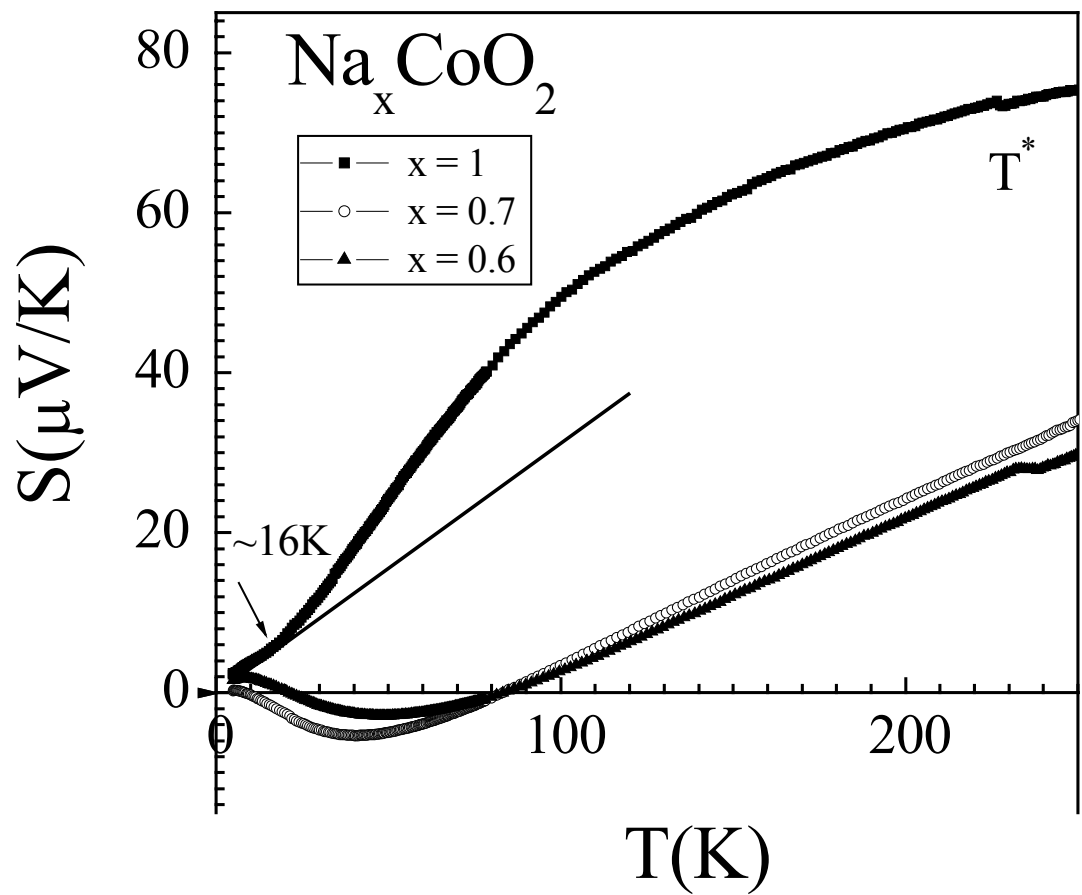